\documentclass[12pt]{article}
\usepackage[dvips]{graphicx}

\usepackage{graphics,epsfig}

\usepackage{psfrag}

\renewcommand{\bar}[1]{\overline{#1}}

\renewcommand{\bar}[1]{\overline{#1}}

\def\ru1{\rule[-0.4truecm]{0mm}{1truecm}}

\begin{document}

\vspace{10mm}

\centerline{\Large \bf 
Transverse Spin Polarization of $\tau^-$}
\vspace*{0.3cm}
\centerline{\Large \bf
in ${\bar{B}}^0\rightarrow D^{+} \tau^- {\bar{\nu}}$ and Charged Higgs Boson}

\vspace{14mm}

\centerline{{\bf Dae Sung Hwang}}

\vspace{7mm}

\centerline{\it Department of Physics, Sejong
University, Seoul 143--747, South Korea}

\vspace{20mm}

\centerline{\bf Abstract}

\vspace{10mm}
\noindent
The spin of the $\tau$ lepton in the semiletonic process
${\bar{B}}^0\rightarrow D^{+} \tau^- {\bar{\nu}}$ can be polarized in
the direction which is normal to the reaction plane,
if the charged Higgs boson exists and its coupling to quarks has a complex phase.
We calculate this transverse polarization of the $\tau$ lepton
by using the experimentally measured $B$ to $D$ transition form factors.
\\

\vfill


\newpage



Since the charged Higg boson effects can be important in the purely leptonic and the semileptonic
decays of $B$ mesons, these decays can be used to extract the constraints on the parameters
of the two-Higgs-doublet model \cite{Hou:1992sy}.
Ref. \cite{Chen:2005gr} showed that the angular asymmetry of the lepton momentum in the semileptonic
$B$-decay is useful for this extraction.
Ref. \cite{Tanaka:2010se} studied the effect of the charged Higgs boson on the longitudinal polarization of
the final tau lepton in the semileptonic $B$-decay.
The transverse polarization of the final tau lepton in the semileptonic $B$-decay was studied
in order to investigate the effect of the charged Higgs boson
\cite{Cheng:1982hq, Atwood:1993ka, Garisto:1994vz, Wu:1997uaa}.
Ref. \cite{Nierste:2008qe} found that we can use the angular distribution of the final pion
produced from the final tau lepton in the semileptonic $B$-decay for this extraction,
and from the experimental result of the branching ratio of $B\to \tau\nu$
they obtained the constraint that the absolute value of $g_S-1$ should be near 1
in the MSSM situation with $g_P=g_S$.
For three example values of $g_S=0, 2$, and $1+i$ which satisfy this constraint,
Ref. \cite{Nierste:2008qe} presented the angular distribution of the final pion.
In this paper we show that the imaginary part of $g_S$ can be extracted by studying the
polarization, transverse to the rection plane, of the final tau lepton in the semileptonic $B$-decay.
These investigations are especially important because of recent impressive experimental progresses
in the processes $B\to D\tau\nu$ and $B\to \tau\nu$
\cite{Bozek:2010xy, Lees:2012xj, Lees:2013uzd, Lees:2012ju, Adachi:2012mm, Kronenbitter:2015kls}. 

{}From Lorentz invariance one finds the decomposition of the
hadronic matrix element
in terms of hadronic form factors:
\begin{eqnarray}
& &<D^+(p)| {\bar c}{\gamma}^{\mu}b |{\bar{B}}^0(P)>
\ =\ (P+p)^\mu f_+(q^2) + (P-p)^\mu f_-(q^2)
\nonumber\\
&=&
\Bigl( (P+p)^\mu
-{M^2-m^2\over q^2}q^\mu \Bigr) \, F_1(q^2)
+{M^2-m^2\over q^2}\, q^\mu \, F_0(q^2) \ .
\label{a1}
\end{eqnarray}
We use the following notations:
$M=m_B$ represents the initial meson mass,
$m=m_D$ the final meson mass,
$m_l$ the lepton mass,
$P=p_B$, $p=p_D$,
and $q_\mu =(P-p)_\mu$.
The form factors $F_1(q^2)$ and $F_0(q^2)$
correspond to $1^-$ and $0^+$ exchanges, respectively.
At $q^2=0$ we have the constraint
$F_1(0)=F_0(0)$,
since the hadronic matrix element in (\ref{a1}) is nonsingular
at this kinematic point.

The differential decay rate is given by
\begin{equation}
d\Gamma ={1\over (2\pi)^3}{1\over 32M^3}|{\cal M}|^2\, dq^2\, dt
\ ,
\label{n1}
\end{equation}
where in the standard model
\begin{equation}
{\cal M}={G_F\over {\sqrt{2}}}\, V_{cb}\, 
<D^+(p)| {\bar c}{\gamma}^{\mu}(1-{\gamma}_5)b |{\bar{B}}^0(P)>\,
{\bar{u_l}}(k)\gamma^\mu (1-\gamma_5)v_\nu({\bar{k}})
\ .
\label{n1z}
\end{equation}
We use the notations: $k=p_{l^-}$, ${\bar{k}}=p_{\bar{\nu}}$,
$q^2=(P-p)^2=(k+{\bar{k}})^2$, $t=(P-{\bar{k}})^2=(p+k)^2$,
and $u=(P-k)^2=(p+{\bar{k}})^2$.

Let us work in the $q$ rest frame.
We use the notation $\Lambda(q^2)={1\over M^2}((M^2+m^2-q^2)^2-4M^2m^2)^{1 / 2}$.
We work in the coordinate system in which we have the following expressions:
$q=({\sqrt{q^2}},0,0,0)$, $P=p_B=(E_B,0,0,|{\vec p}_D|)$, $p=p_D=(E_D,0,0,|{\vec p}_D|)$,
$p_l=(E_l,|{\vec p}_l|\sin{\theta},0,|{\vec p}_l|\cos{\theta})$,
$|{\vec p}_D|={M^2\over 2{\sqrt{q^2}}}\Lambda(q^2)$, $E_B={1\over 2{\sqrt{q^2}}}(M^2-m^2+q^2)$,
$E_D={1\over 2{\sqrt{q^2}}}(M^2-m^2-q^2)$, $|{\vec p}_l|={1\over 2{\sqrt{q^2}}}(q^2-m_l^2)$,
$E_l={1\over 2{\sqrt{q^2}}}(q^2+m_l^2)$.

From (\ref{a1}), (\ref{n1}) and (\ref{n1z}), we get \cite{koerner,Hwang:2000xe}
\begin{eqnarray}
{d\Gamma ({\bar{B}}^0\rightarrow D^+ l^- {\bar{\nu}})\over dq^2d\cos{\theta}}&=&
{G_F^2\over 2^8\pi^3}\, |V_{cb}|^2\, M^3\, \Lambda(q^2)\,
(1-{m_l^2\over q^2})^2
\label{aa3z}\\
&\times&
\Big[\, -(\Lambda(q^2))^2\, (1-{m_l^2\over q^2})\, F_1(q^2)^2\, {\rm cos}^2{\theta}
\nonumber\\
&&
\, +\,
2\, \Lambda(q^2)\, {m_l^2\over q^2}\, (1-{m^2\over M^2})\,
F_1(q^2)\, F_0(q^2)\, \cos{\theta}
\nonumber\\
&&
\, +\, (\Lambda(q^2))^2\, F_1(q^2)^2 \, +\, {m_l^2\over q^2}\, (1-{m^2\over M^2})^2\,
F_0(q^2)^2\, \Big]\, ,
\nonumber
\end{eqnarray}
where the allowed range of $q^2$ is given by
\begin{equation}
m_l^2\le q^2\le (M-m)^2.
\label{aa3a}
\end{equation}
After the $\theta$ integration of (\ref{aa3z}) over the range $0\le \theta \le \pi$,
the $q^2$ distribution of the decay rate is given by
\begin{eqnarray}
&&{d\Gamma ({\bar{B}}^0\rightarrow D^+ l^- {\bar{\nu}})\over dq^2}=
{G_F^2\over 192\pi^3}\, |V_{cb}|^2\, M^3\, \Lambda(q^2)\,
(1-{m_l^2\over q^2})^2\times
\label{aa3}\\
&&
\Big[\, (\Lambda(q^2))^2\, (1+{1\over 2}{m_l^2\over q^2})\, F_1(q^2)^2
\, +\, {3\over 2}\, {m_l^2\over q^2}\, (1-{m^2\over M^2})^2\,
F_0(q^2)^2\, \Big]\, .
\nonumber
\end{eqnarray}

When there exists the charged Higgs boson,
the transition matrix element ${\cal M}$ given in (\ref{n1z}) is modified
by the contribution of the charged Higgs boson:
\begin{eqnarray}
{\widetilde{\cal M}}&=&{G_F\over {\sqrt{2}}}\, V_{cb}\,\Big[\,  
<D^+(p)| {\bar c}{\gamma}^{\mu}(1-{\gamma}_5)b |{\bar{B}}^0(P)>\,
{\bar{u_l}}(k)\gamma^\mu (1-\gamma_5)v_\nu({\bar{k}})
\label{n1zz}\\
&&-\
{{\bar m_b}m_l\over M^2}\, 
<D^+(p)| {\bar c} (g_S+g_P{\gamma}_5) b |{\bar{B}}^0(P)>\,
{\bar{u_l}}(k) (1-\gamma_5) v_\nu({\bar{k}})
\, \Big]\ .
\nonumber
\end{eqnarray}
Using
\begin{equation}
<D^+(p)| {\bar c}b |{\bar{B}}^0(P)>
\ =\ {M^2-m^2\over {\bar m_b}-{\bar m_c}}\, F_0(q^2) \ ,
\label{a1zzz}
\end{equation}
from (\ref{a1}), (\ref{n1}) and (\ref{n1zz}), we get
\begin{eqnarray}
{d\Gamma ({\bar{B}}^0\rightarrow D^+ l^- {\bar{\nu}})\over dq^2d\cos{\theta}}&=&
{G_F^2\over 2^8\pi^3}\, |V_{cb}|^2\, M^3\, \Lambda(q^2)\,
(1-{m_l^2\over q^2})^2
\label{aplusb}\\ 
&&\times\ \Big[\, {1\over 2}\, A(q^2,\theta )\ +\ {1\over 2}\, B(q^2,\theta )\,\,
{\vec n}\cdot {\hat y}\, \Big]\ ,
\nonumber
\end{eqnarray}
where
${\hat y}=({\vec p_D}\times {\vec p_l}) / |{\vec p_D}\times {\vec p_l}|$,
${\vec n}$ is unit vector in the spin direction of $l^-$, and
\begin{eqnarray}
A(q^2,\theta )&=&
\Big[\, -(\Lambda(q^2))^2\, (1-{m_l^2\over q^2})\, F_1(q^2)^2\, {\rm cos}^2{\theta}
\label{Aterm}\\
&&
\, +\,
2\, \Lambda(q^2)\, {m_l^2\over q^2}\, (1-{m^2\over M^2})\,
{\rm Re}[F_1(q^2)\, {\widetilde F}_0(q^2)]\, \cos{\theta}
\nonumber\\
&&
\, +\, (\Lambda(q^2))^2\, F_1(q^2)^2 \, +\, {m_l^2\over q^2}\, (1-{m^2\over M^2})^2\,
|{\widetilde F}_0(q^2)|^2\, \Big]\ ,
\nonumber\\
B(q^2,\theta )&=&
-2\, {\rm Im}[g_S]\, {1\over 1-{{\bar m_c}\over {\bar m_b}}}\,
\Lambda(q^2)\, (1-{m^2\over M^2})\,
{m_l\over M}\, {{\sqrt{q^2}}\over M}\,
F_1(q^2)\,  F_0(q^2)
\sin{\theta}\ ,
\label{Bterm}
\end{eqnarray}
where
\begin{equation}
{\widetilde F}_0(q^2)\, =\, 
\Big( \, 1- g_S\, {1\over 1-{{\bar m_c}\over {\bar m_b}}}\, {q^2\over M^2}\, \Big)\, F_0(q^2)\ .
\label{gstilda}
\end{equation}
In (\ref{Bterm}) we used the notation
$g_S={\rm Re}[g_S]+i\, {\rm Im}[g_S]$ in which ${\rm Re}[g_S]$ and ${\rm Im}[g_S]$ are real.
The second term in (\ref{aplusb}) is proportional to the vector product
${\vec n}\cdot ({\vec p_D}\times {\vec p_l})$ which is the correlation of the
$l^-$ spin and the momenta of $D^+$ and $l^-$, and so it corresponds to a single-spin asymmetry.
From (\ref{aplusb}) we obtain the transverse spin polarization of $l^-$ as
\begin{equation}
{\cal P}_y(q^2,\theta )=
{d\Gamma [{\vec n}\cdot {\hat y}=+1] - d\Gamma [{\vec n}\cdot {\hat y}=-1]\over
d\Gamma [{\vec n}\cdot {\hat y}=+1] + d\Gamma [{\vec n}\cdot {\hat y}=-1]}
={B(q^2,\theta )\over A(q^2,\theta )}\ .
\label{PolarizationFormula}
\end{equation}

For the $B$ to $D$ meson (heavy to heavy) transition form factors, the heavy
quark effective theory gives \cite{IW8990}
\begin{equation}
F_1(q^2)={M+m\over 2{\sqrt{M m}}}\, {\cal G}(\omega)\ , \qquad
F_0(q^2)={2{\sqrt{M m}}\over M+m}\, {\omega + 1 \over 2}\,
{\cal G}(\omega)\ ,
\label{ff1}
\end{equation}
where $\omega = (M^2 + m^2 - q^2)/(2M m)$ and
${\cal G}(\omega)$ is a form factor which becomes the Isgur-Wise function in
the infinite heavy quark mass limit.
We use the parameterization of ${\cal G}(\omega)$
given in \cite{Caprini}:
\begin{equation}
{{\cal G}(\omega)\over {\cal G}(1)} =
1-8{\rho}_{D}^2 z + (51 {\rho}_{D}^2 - 10 )z^2 -
(252 {\rho}_{D}^2 - 84 ) z^3\ 
\label{ff2}
\end{equation}
with
$z = {({\sqrt{\omega +1}} - {\sqrt{2}}) / ({\sqrt{\omega +1}} + {\sqrt{2}})}$.
We use the world average values given in \cite{Luth:2011zz}:
\begin{equation}
{\cal G}(1)|V_{cb}|\times 10^3=42.69 \pm 0.72 \pm 1.27\ , \qquad
{\rho}_{D}^2=1.20 \pm 0.04 \pm 0.04 \ .
\label{ff4}
\end{equation}

For explicit calculations with (\ref{aplusb}) and (\ref{PolarizationFormula}),
we use the form factors given by (\ref{ff1}), (\ref{ff2}) and (\ref{ff4}).
Ref. \cite{Nierste:2008qe} obtained
the constraint that the absolute value of $g_S-1$ should be near 1
in the MSSM situation with $g_P=g_S$ from
the experimental result of the branching ratio of $B\to \tau\nu$.
We perform explicit calculations
at first with three example values of $g_S=0, 2$, and $1+i$ which satisfy this constraint,
following what was done in Ref. \cite{Nierste:2008qe}.
Then, we also consider a more general case of $g_S-1=e^{i\alpha}$, which satisfies
the above-mentioned constraint.
For the numerical value of ${{\bar m_c}\over {\bar m_b}}$ appearing in (\ref{Bterm}),
we use $0.2$ which is the value Ref. \cite{Nierste:2008qe} used.

\begin{figure}
\centering
\begin{minipage}[t]{10.0cm}
\centering
\includegraphics[width=\textwidth]{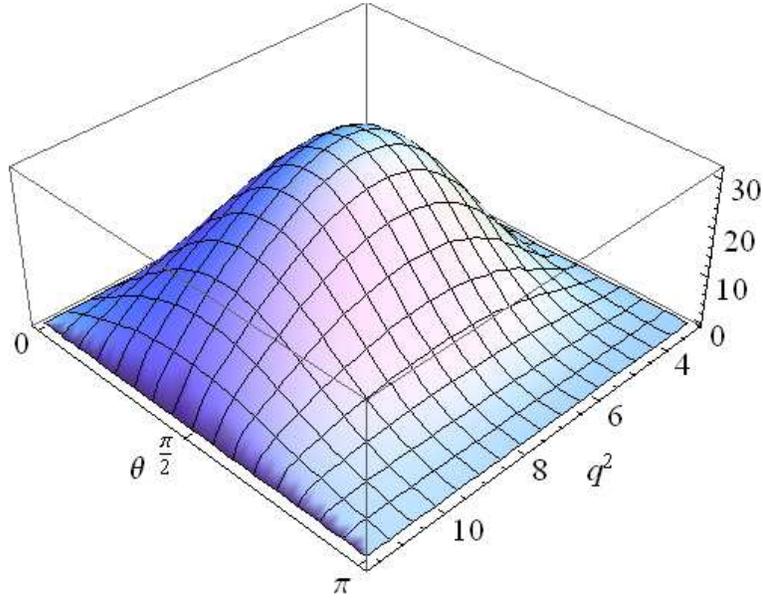}
\end{minipage}\hspace{1.0cm}
\parbox{0.95\textwidth}{\caption{
\it $10^{17}\times {d\Gamma / dq^2\, d\theta}\ ({\rm GeV}^{-1})$ for $g_S=0$.
\label{fig1}}}
\end{figure}

\begin{figure}
\centering
\begin{minipage}[t]{10.0cm}
\centering
\includegraphics[width=\textwidth]{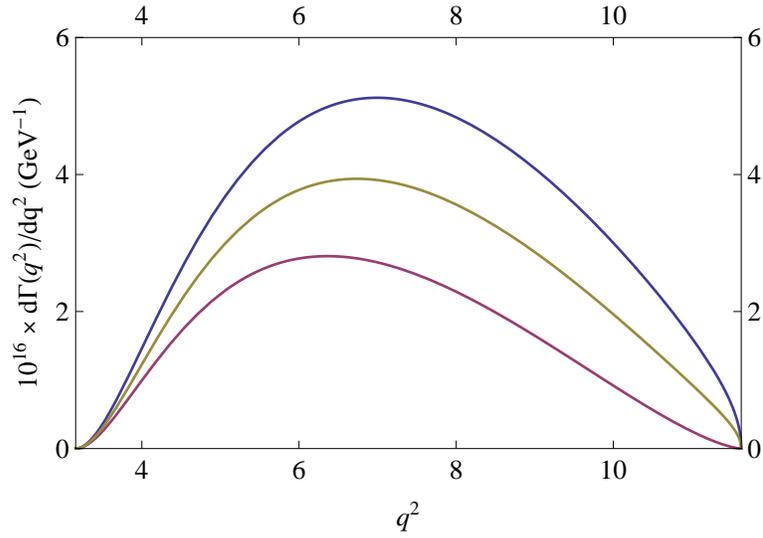}
\end{minipage}\hspace{1.0cm}
\parbox{0.95\textwidth}{\caption{
\it $10^{16}\times {d\Gamma / dq^2}\ ({\rm GeV}^{-1})$ for $g_S=0, \ 1+i, \  2$
in sequence from the top to the bottom.
\label{fig2}}}
\end{figure}

\begin{figure}
\centering
\begin{minipage}[t]{10.0cm}
\centering
\includegraphics[width=\textwidth]{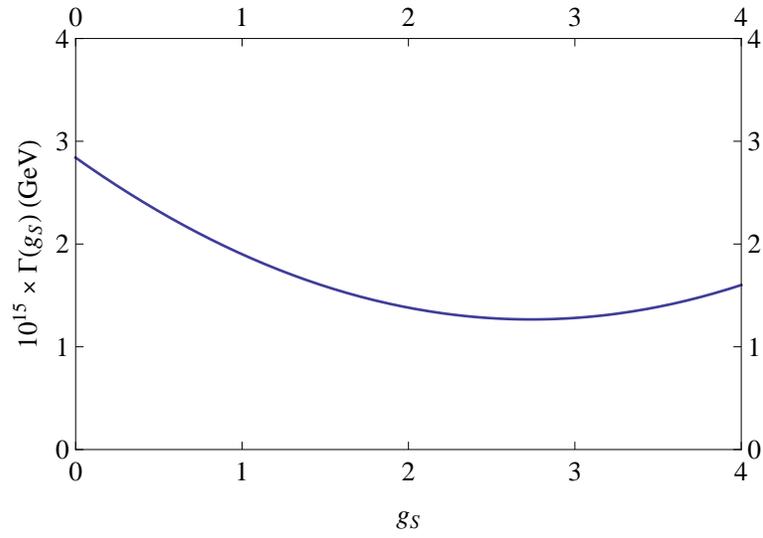}
\end{minipage}\hspace{1.0cm}
\parbox{0.95\textwidth}{\caption{
\it $10^{15}\times \Gamma (g_S)\ ({\rm GeV})$ for real $g_S$
\label{fig3}}}
\end{figure}

\begin{figure}
\centering
\begin{minipage}[t]{10.0cm}
\centering
\includegraphics[width=\textwidth]{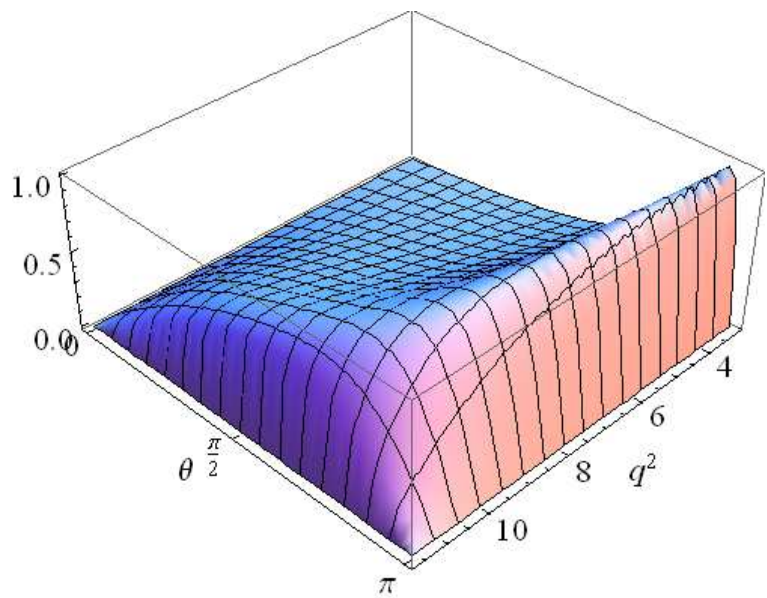}
\end{minipage}\hspace{1.0cm}
\parbox{0.95\textwidth}{\caption{
\it ${\cal P}_y(q^2,\theta)$ for $g_S=1+i$.
\label{fig4}}}
\end{figure}

\begin{figure}
\centering
\begin{minipage}[t]{10.0cm}
\centering
\includegraphics[width=\textwidth]{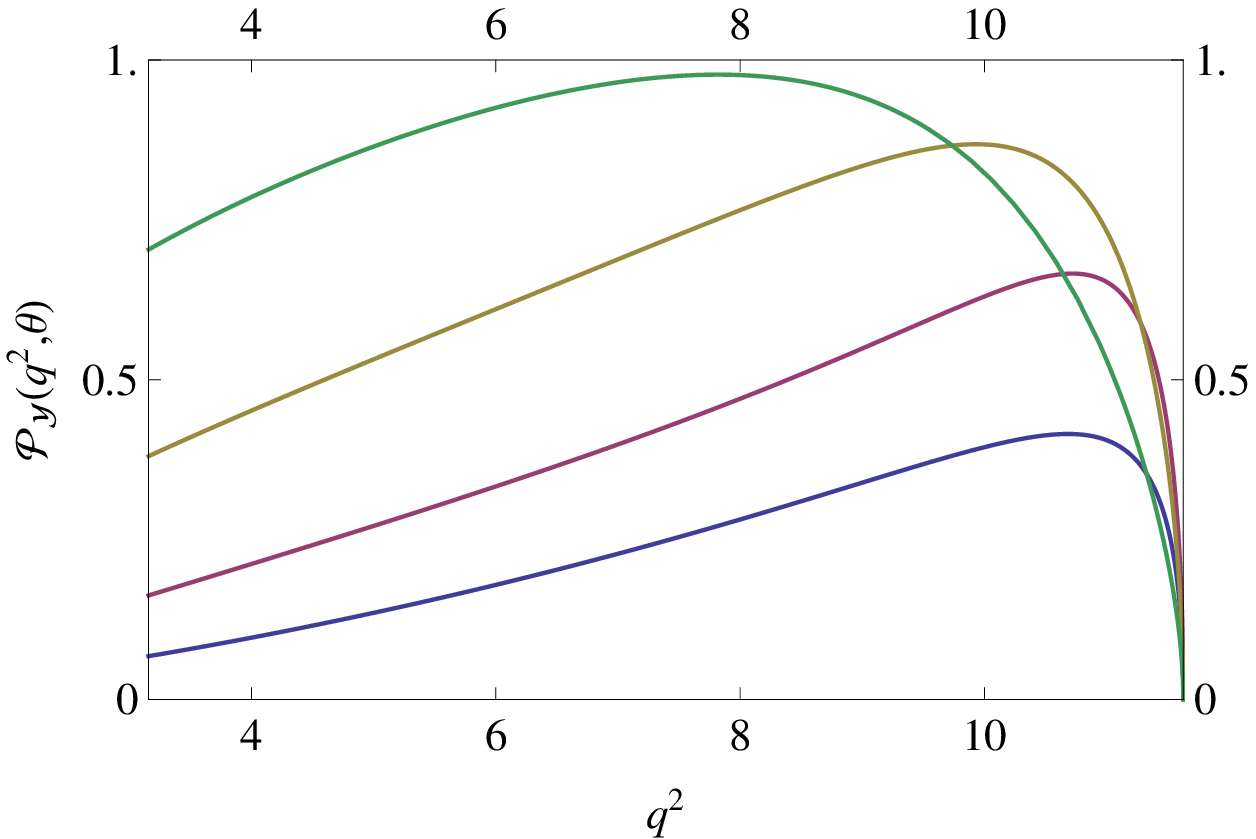}
\end{minipage}\hspace{1.0cm}
\parbox{0.95\textwidth}{\caption{
\it ${\cal P}_y(q^2,\theta)$ for $g_S=1+i$,
at $\theta ={\pi\over 4},\ {\pi\over 2},\ 
{3\pi\over 4},\ {7\pi\over 8}$ in sequence from the bottom to the top.
\label{fig5}}}
\end{figure}

\begin{figure}
\centering
\begin{minipage}[t]{10.0cm}
\centering
\includegraphics[width=\textwidth]{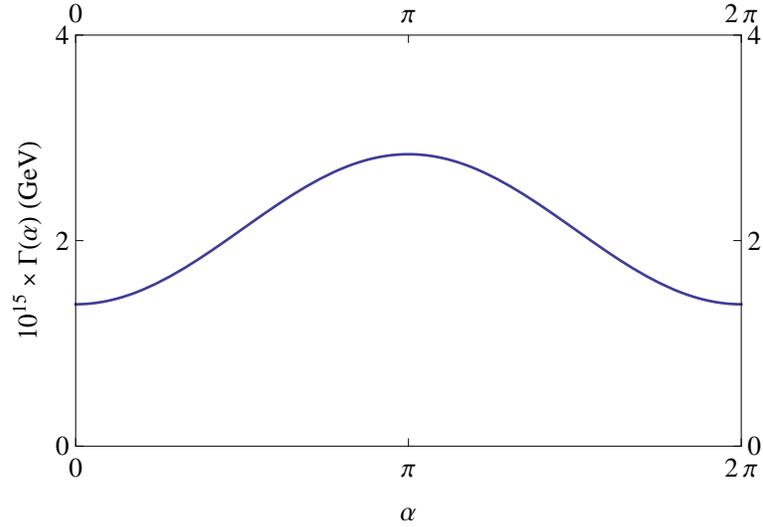}
\end{minipage}\hspace{1.0cm}
\parbox{0.95\textwidth}{\caption{
\it $10^{15}\times\Gamma(\alpha)$ (GeV),
$\alpha$ dependence of the decay rate.
\label{fig6}}}
\end{figure}

\begin{figure}
\centering
\begin{minipage}[t]{10.0cm}
\centering
\includegraphics[width=\textwidth]{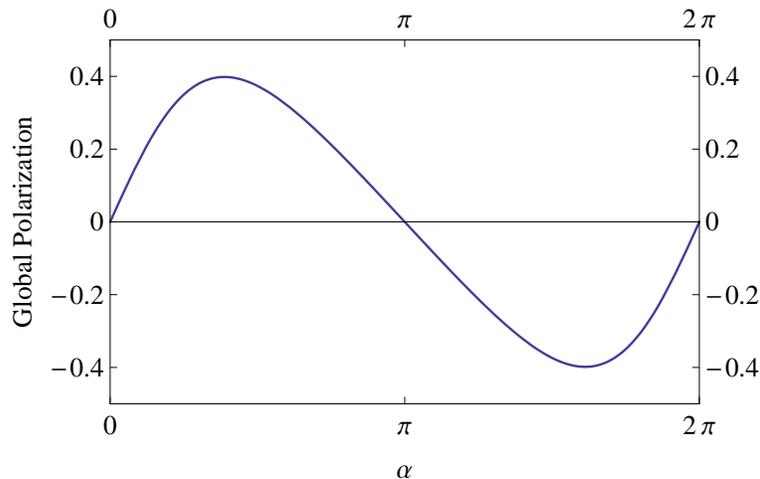}
\end{minipage}\hspace{1.0cm}
\parbox{0.95\textwidth}{\caption{
\it $\alpha$ dependence of the global polarization.
\label{fig7}}}
\end{figure}

Fig. \ref{fig1} shows the result from (\ref{aplusb}) for $g_S=0$,
which is the same as that from (\ref{aa3z}).
By integrating (\ref{aplusb}) over the angle, we obtain the graphs given in
Fig. \ref{fig2} for $g_S=0, \ 1+i, \  2$.
For reference, we calculate the decay rate for real values of $g_S$ in the range of
$0\le g_S \le 4$ by integrating (\ref{aplusb}) over the angle and $q^2$
to get the result presented in Fig. \ref{fig3}.
We calculate the transverse spin polarization of the tau lepton by using
(\ref{PolarizationFormula}) when $g_S=1+i$, and its result is given in Fig. \ref{fig4}.
The graphs in Fig. \ref{fig5} are given from the result presented in Fig. \ref{fig4} at
fixed values of $\theta$;
$\theta ={\pi\over 4},\ {\pi\over 2},\ {3\pi\over 4},\ {7\pi\over 8}$.
Then, we consider the case of $g_S-1=e^{i\alpha}$
by adopting the constraint obtained in Ref. \cite{Nierste:2008qe}.
We calculate the decay rate by integrating (\ref{aplusb}) over the angle and $q^2$
and obtain the $\alpha$ dependence of the decay rate presented in Fig. \ref{fig6}.
We also calculate the global polarization, which is defined as
the integration of the numerator in (\ref{PolarizationFormula}) over the angle and $q^2$
divided by that of the denominator in (\ref{PolarizationFormula}),
and get the result given in Fig. \ref{fig7}.

We derived the formula for the transverse spin polarization of
the $\tau$ lepton in the semiletonic process
${\bar{B}}^0\rightarrow D^{+} \tau^- {\bar{\nu}}$,
which is expressed in terms of the coupling constant $g_S$ and
the $B$ to $D$ transition form factors.
This formula shows that the $\tau$ lepton can be transversely polarized
when the charged Higgs boson exists and its coupling to quarks has a complex phase.
We performed explicit calculations of this transverse polarization
by using the experimentally measured $B$ to $D$ transition form factors.
Experimental investigation of this transverse polarization, with the BaBar and Belle data
and in LHCb and Belle II, should be important for the study of the charged Higgs boson.
If this transverse polarization is measured to be nonzero experimentally, it could be the
implication that the charged Higgs boson exists and its coupling to quarks has a complex phase.

\vfill\pagebreak


\noindent
{\em Acknowledgements} \\
\noindent
This work was supported in part
by the Korea Foundation for International
Cooperation of Science \& Technology (KICOS)
and the Basic Science Research Programme through the
National Research Foundation of Korea (2010-0011034).
\\



\end{document}